\documentstyle[prl,aps]{revtex}

\begin{document}
\twocolumn[\hsize\textwidth\columnwidth\hsize\csname@twocolumnfalse%
\endcsname
\title{The Ground State of the ``Frozen'' Electron Phase in Two-Dimensional
Narrow-Band Conductors with a Long-Range Interelectron Repulsion.
Stripe Formation and Effective Lowering of Dimension.}
\author{A.~A.~Slutskin, V.~V.~Slavin, and H.~A.~Kovtun.}
\address{Institute for Low Temperature Physics and  Engineering, 47 Lenin
Ave., Kharkov, Ukraine.}
\date{March,6, 1999}
\maketitle
\draft
\begin{abstract}

In narrow-band conductors a weakly screened  Coulomb interelectron
repulsion can supress  narrow-band electrons' hopping, resulting in
formation of a ``frozen'' electron phase which differs principally from
any known macroscopic self-localized electron state including the Wigner
crystal.  In a zero-bandwidth limit the ``frozen''  electron phase is a
classical lattice system with a long-range interparticle repulsion. The
ground state of such systems has been considered in the case of {\em two}
dimensions for an {\em isotropic} pair potential of the mutual  particle
repulsion. It has been shown that particle ordering into stripes and
effective lowering of dimension universally resides in the ground state
for any physically reasonable pair potential and for any geometry of the
conductor lattice.  On the basis of this fact a rigorous general procedure
to fully describe  the ground state has been formulated .
Arguments have been adduced that charge ordering in High-$T_c$
superconductors \cite{HighTc1,HighTc2} testifies to presence of a
``frozen'' electron phase in these systems.

\end{abstract}
 ]

\subsection{Introduction.}
\label{a}

High-T$_c$ superconductors studies have caused a surge of interest in
properties of narrow-band layered and two-dimensional (2D) conductors. An
important consequence of the layerness is substantial weakening of the
screening of a Coulomb interaction between the charge carriers. (The
screening radius cannot, under any circumstances, be less than the
interlayer distance).  Besides, in layered conductors it is possible
to well separate the charge carriers (for the sake of definiteness, we
consider them electrons) from the donors, so that the mean energy,
$u_{ee}$, of the long-ranged interelectron repulsion prevails over the
energy of an electron attraction to the donors. Under these conditions it
is the mutual repulsion of narrow-band electrons that can {\em supress}
their tunneling between the equivalent orbits of the conductor lattice,
resulting in formation of a ``frozen'' electron phase (FEP) which differs
principally from any known macroscopic self-localized electron state
including the Wigner crystal \cite{FNT93_UFN95}. The FEP occurs when the
electron bandwidth, $t$, is less than $\delta u = (a/ r_{ee})u_{ee}$,
where $\delta u$ is the typical change in the narrow-band electron Coulomb
energy in electron hopping, $a$ is the range of hopping,
$r_{ee}$ is the mean electron separation.  The high-$T_c$ cuprates, grain
boundaries of polycrystal electroceramic materials \cite{Heywang}, as well
as some art 2D conductors \cite{Pepper,Nejoh,Osifchin} appear to be most
favorable for 2D FEP coming to existence.

Macroscopical behavior of the 2D FEP is rather unconventional.  Its
distinctive features are rooted in properties of its ground state (GS) at
$t\ll \delta u$.  In the limit $t/\delta u \rightarrow 0$ the GS of the 2D
FEP is much the same as that of other 2D lattice systems with a
long-ranged interparticle repulsion. (An example is an ensemble of
adsorbed atoms strongly interacting with their substrate and mutually
repelling each other \cite{Pokrovsky}). As far as we know, neither the
thermodynamics nor the GS of such systems have been studied adequately.
 Here we offer a unified approach to the description of the GS of the 2D
zero-bandwidth FEP (and similar lattice systems) with an {\em isotropic}
pair potential of the interelectron repulsion, $v(r)$ ($r$ is the distance
between interacting electrons).  The key point of our consideration is a
new phenomenon --- a {\em zero-temperature effective lowering of dimension
(LOD)} --- which we have revealed to underlay ({\em despite the pair
potential isotropy}) the main GS properties of the 2D FEP for: i/
arbitrary arrangement of the sites which can be  occupied by electrons
provided the sites constitute a primitive lattice (it is called host
lattice below); ii/any filling factor, $\rho = N/{\cal N}$ ($N\rightarrow
\infty$ and $\cal N\rightarrow \infty$ are the total numbers of the
electrons and the host-lattice sites respectively); iii/ any physically
reasonable $v(r)> 0$. We take the term LOD to mean that the GS of the 2D
FEP is a set of different effective 1D FEP whose ``particles'' are
periodic stripes on the lattice od the 2D conductor. For each 1D system
of the set there is its own $\rho$ interval where this 1D FEP represents
the 2D one, the whole range, $0\leq \rho\leq 1$, comprising all the
intervals. The LOD enables us to offer a rigorous analytical procedure
for the 2D FEP GS description, using the exact results of the general
theory of the 1D lattice systems with a long-ranged interparticle
repulsion\cite{Hubbard,Bak,Synay}.

\subsection{ Hamiltonian. S-crystals.}
\label{b}
The Hamiltonian, $\cal H$, of the system under consideration has the form

\begin{equation}
\label{H}
{\cal H}\{n(\vec r)\} ={\frac12}\sum_{\vec r\neq\vec {r'}}
{v\left(|\vec r-\vec {r'}|\right)n(\vec r)n(\vec {r'})}\,,
\end{equation}

\noindent where $\vec r= m_1\vec a_1 + m_2\vec a_2$ are radius vectors of
the host-lattice sites, $m_{1,2}$ are integers, $\vec a_{1,2}$ are
host-lattice primitive translation vectors (PTVs); the occupation numbers
of the host-latice sites, $n(\vec r)=0 \text{ or } 1$, are microscopic
variables \cite{footnote1}; the sum is taken over the whole host lattice.
The pair potential is assumed to be an everywhere convex function of the
form $v(r) = \tilde v(r)/r$, where function $\tilde v(r)$ depends on the
character of the screening medium and its position with respect to the 2D
FEP. In any case $\tilde v(r)$ tends to zero as $r^{-2}$ or faster when $r
\rightarrow \infty$; $\tilde v(0) = e^2/\kappa$ ($e$ is the electron
charge, $\kappa$ is the dielectric permittivity). Otherwise $\tilde v(r)$
can be reckoned as arbitrary: as will be shown below, its specific form is
immaterial to our approach.

Among the GS configurations $\{n(\vec r)\}$ with different $\rho$ the
{\em simplest} ones are 2D crystals with one electron per cell
(``S-crystals'').  Their inverse $\rho$ values make up an infinite set of
integers $Q_j = |\det (m^j_{\kappa\lambda}(\vec a_1,\vec a_2))|$, where
$j$ indexes S-crystals, integers $m^j_{\kappa\lambda}$ ($\kappa,\lambda
=1,2$) are components of S-crystal PTVs in the $\vec a_\lambda$ basis;
$Q_j$ is the $j$-th S-crystal elementary-cell area measured in units of
that of the host lattice, $\sigma_0 = |\vec a_1\times \vec a_2|$.

Our strategy is to derive the full description of the GS for any $\vec
a_{1,2}$, starting with consideration of small vicinities of $\rho=1/Q_j$.
Since specific $m^j_{\kappa\lambda}$ values are irrelevant to this
reasoning, we drop index $j$ at $Q$ and at other characteristics of the
S-crystals for a while.

Due to discreteness of the system with the Hamiltonian (\ref{H}) a
macroscopically small change, $\delta \rho$, in $\rho$ ($\delta
\rho\rightarrow 0, \;N^{1/2}|\delta \rho|\rightarrow\infty$ when
$N\rightarrow\infty$) produces only {\em isolated defects} in an
S-crystal, the space structure of the defects essentially depending on
whether they result from an increase or a decrease in $\rho$. This fact
is expressed by the identity

\begin{equation}
\label{identity}
\begin{array}{c}
E_g(N\pm |\delta N|,{\cal N}\pm
|\delta {\cal N}|) - E_g(N,{\cal N}) =\\
\pm\mu_{\pm}|\delta N|\mp P_{\pm}|\delta {\cal N}|,
\end{array}
\end{equation}

\noindent where $E_g$ is the GS energy, $\delta N$ and $\delta {\cal N}$
are changes in $N$ and $\cal N$ producing $\delta \rho$. The
proportionality coefficients, $\mu_- < \mu_+$, $P_- < P_+$, are the values
of the chemical potential, $\mu$, and the pressure, $P$, which are the
endpoints of the $\mu$ and $P$ intervals of S-crystal existence. They are
determined by the energies of formation of corresponding defects.  Thus,
in some vicinity of $\rho = 1/Q$ the GS is bound to be a superstructure of
the defects.  Our next step is to find them.

\subsection{Zero-dimensional defects and their coalescence.}
\label{defecton} Adding to or removing from an S-crystal {\em one}
electron results in formation of a zero-dimensional defect,
``$+$defecton'' or ``$-$defecton'' respectively. One can be inclined to
think that $\delta N$ should be identified exactly with the total number
of $\pm$defectons spatially separated, $\pm\mu_{\pm}$ being simply the
energy of $\pm$defecton formation, $\epsilon_\pm$.  However, this
seemingly evident statement is actually incorrect due to a {\em
coalescence} of defectons of the same ``sign''. In other words, if the
number, $|\nu|$, of S-crystal electrons removed ($\nu < 0$) or added ($\nu
>0$) is more than $1$, a {\em bound state} of $|\nu|$ $\pm$defectons
arises whose energy is less than $|\nu|\,\epsilon_\pm$. We have revealed
the coalescence by computation, using a "dipole" description of the GS
with $\nu = \pm 1, \pm2,\ldots$, which we have specially worked out for
this purpose.  The dipole approach offers a clear view of how the
defectons' bound state arises despite the fact that the defectons of the
same sign repel each other, being widely spaced.

At $\nu \neq 0$ a perturbed S-crystal is formed where beside electrons
placed at host-lattice sites in the interstices of the S-crystal ($\nu>
0$) or empty S-crystal sites, ``holes'', ($\nu< 0$) there are generally a
certain number of S-crystal electrons shifted from their native S-crystal
sites.  The latter can be considered as ``antiparticles'' whose charge is
equal to the electron one in magnitude but is opposite in sign, a pair ``a
electron shifted by a vector $\vec \xi$ + its antiparticle located at an
S-crystal site $\vec r$~'' being the ``$\vec r,\vec \xi$-dipole''.  Thus,
the perturbation of the S-crystal can be envisioned as an ensemble
consisting of several dipoles and $|\nu|$ interstitial particles/holes
(IP/Hs). The dipoles interact with the IP/Hs and with each other. The
energy of interaction between the IP/H (at $\vec r= 0$) and $\vec r,\vec
\xi$-dipole is $u_{\vec \xi}\,(\vec r) = {\text {sign}}\,\nu\,(v(|\vec r
-\vec \xi|) - v(|\vec r|)) \equiv {\text {sign}}\,\nu \widehat\Delta_{\vec
\xi}\, v(|\vec r|)$; the energy of interaction between $\vec r,\vec \xi$-
and $\vec r^{\,\prime},\vec \xi^{\,\prime}$-dipole is $u_{\vec \xi,\vec
\xi^{\,\prime}}(\vec r - \vec r^{\,\prime})= \widehat\Delta_{\vec
\xi}\,\widehat\Delta_{\vec \xi^{\,\prime}}v(|\vec r - \vec
r^{\,\prime}|)$.  The IP/Hs, in turn, undergo a mutual repulsion and are
exposed to an ``external'' field, $u(\vec r)$, which is equal to $- 2u_0$
for holes (here and further on $u_{ee}$ of the S-crystal is denoted by
$u_0$), and for IPs it is the field produced at a point $\vec r$ by all
 electrons of the S-crystal.  In these terms the change in the GS energy
at a given $\nu$, $U_{\text {GS}}(\nu)$, takes the form

\begin{equation}
\label{dipole}
U_{\text {GS}}(\nu)=
\min\left(V_{\text {rep}} + U_d +  U_{\text {exc}} +
U\right).
\end{equation}

\noindent Here $V_{\text {rep}} =\sum_{\alpha<\beta} v(|\vec
r_{\alpha\beta}|)$ is the energy of the mutual repulsion of the IP/Hs;
$U_d= \sum_{\alpha,i}u_{\vec \xi_i}\,(\vec r_{\alpha i})$ is the energy of
their interaction with the dipoles; $U_{\text {exc}} = \sum_i\delta
u_{\vec \xi_i} + \sum_{i<k}u_{\vec \xi_i,\vec \xi_k}(\vec r_{i k})>0$ is
the excitation energy of S-crystal with $n_d$ dipoles at $\nu= 0$; $\delta
u_{\vec \xi} \sim u_0|\vec \xi|^2/r_{ee}^2 > 0$ is the energy of formation
of one dipole; $U =\sum_\alpha u(\vec r_\alpha)$ is the energy of the
IP/Hs in the external field mentioned; indexes $\alpha = 1,\ldots, |\nu|$
and $i=1,\ldots,n_d$ enumerate the IP/H raduis-vectors and dipoles
respectively, $n_d$ is the total number of the dipoles; $\vec r_{ab}\equiv
\vec r_a -\vec r_b$.  The minimum is taken in respect to $n_d$, the dipole
variables, $\vec r_i, \vec \xi_i$, and $\vec r_\alpha$.  Therefore, the
dipole approach allows to work with only a few discrete variables. This
facilitates considerably the Monte-Carlo computer simulation of the
$\pm$defectons ($U_{\text {GS}}(\pm 1)= \pm\epsilon^\pm$) and their
coalescence at $|\nu| >1$.

The mechanism of the coalescence can be elucidated by the following
heuristic arguments. The GS total dipole energy, $E_d(\nu)= U_{\text
{exc}}(\nu) + U_d(\nu)$, is negative, so that for any $|\nu|$ the GS space
structure is determined by an interplay between negative $U_d$ and
positive $U_{\text {exc}}\:$, $V_{\text {rep}}$.  The IP/H -- dipole
interaction gives the maximal gain in energy when {\em each} IP/H is
embedded in a ``shell'' of four dipoles which are attracted to it, the
dipoles' antiparticles forming a parallelogram of a size $\sim r_{ee}\sim
Q^{1/2}$ (Fig.1).  The shells of neighboring IP/Hs are bound to share some
of their dipoles for $U_{\text {exc}}$ (and hence $n_d$) to be as small as
possible. This requirement can be fulfilled only when all IP/Hs are {\em
aligned in a row}, the near-neighbor IP/Hs being shifted relative to one
another by the same S-crystal PTV with the modulus $\sim r_{ee}$.
(Fig.1).  In such a case $|E_d(\nu)|$ is more than the magnitude of the
dipole energy of $|\nu|$ infinitely separated defectons, $E_d^\infty=|\nu|
E_d(\pm 1)$.  The coalescence arises when the energy gain, $\Delta
= |E_d(\nu)| - |E_d^\infty|$, exceeds $V_{\text {rep}}$ of the IP/Hs
aligned in the row. Since $\Delta \sim |\nu|v(r_{ee})$ this condition is
met if $v(r)$ decreases not too slowly, or, more exactly, if

\begin{equation}
\label{gamma}
\gamma
= \int_{r_{ee}}^\infty\limits  v(r)dr/r_{ee} v(r_{ee})\lesssim 1.
\end{equation}

\noindent The computer simulation carried out with the model potential
$v(r) \propto r^{-\beta}\exp(-r/R)$ over a wide range of the parameters,
$\beta,R $, has confirmed that the condition (\ref{gamma}) is really the
criterion of the coalescence for any $|\nu|$ (and any $\vec a_{1,2}$).

Criterion \ref{gamma} is for the most part fulfilled.  It holds for any
$\tilde v(r)$ (section \ref{b}) such that $\tilde v(0) -\tilde v(r_{ee})
\sim \tilde v(0)$. This case will be the focus of our attention
from here on. Parameter $\gamma$ becomes $\gg 1$ if $\tilde v(r)$
decreases substantially only for $r$ which are exponentially large in
$\gamma$.  In this limit the mutual repulsion of the IP/Hs disrupts their
row, and there is no coalescence, at least for sufficiently large
$|\nu|$.  However, in section \ref{bigR} it is outlined that the LOD
governs the GS in this rather special case, too.

\subsection{The lowering of dimension.}
\label{d}
\paragraph{The elementary stripes in the 2D FEP.}
\label{1D}
As follows from the aforesaid, the bound state of $|\nu|$ defectons is
transformed into a periodic {\em stripe-like} structure with an infinite
increase in $|\nu|$. (Fig.~1).  It consists of elementary 1D defects
which, as will be shown below, repel each other.  Therefore, it is {\em
the simplest 1D defects} that are expected to form the GS superstructure.
An arbitrary 1D defect of such a type is a stripe of rarefaction or
compression which arises when an S-crystal part adjacent to a line of
electrons with some PTV, $\vec d$, is shifted as a whole relative to the
other one by a host-lattice translation vector, $\vec \xi$.  Formation of
one stripe of length $L_s$ changes $\cal N$ by $\delta {\cal N} =\pm
\sigma L_s$ ($\sigma= |\vec d\times \vec \xi|$, $L_s$ is measured in units
of $|\vec d\,|$ ). The corresponding change in energy, $\delta E$, is
proportional to $\delta {\cal N}$:

\begin{equation}
\label{dE}
\delta E/|\delta {\cal N}| = \varepsilon(\vec d,\vec \xi)\, =
\sigma^{-1}\sum\nolimits_{n=1}^{\infty}{\sum\nolimits_{\vec
r}}^{\prime}u_{\vec\xi}\,(\vec r - n\vec f).
\end{equation}

\noindent Here ${\sum_{\vec r}}^{\prime}$ means summation over the
S-crystal semiplane $\vec r = k\vec d + l\vec f$ ($-\infty < k < \infty$,
$-\infty< l\leq 0 $); $\vec f$ is any S-crystal PTV other than $\vec d$.
The GS is realized by the stripes with $\vec d = \vec d_\pm$ and $\vec \xi
= \vec \xi_{\pm}$ which minimize $\varepsilon(\vec d,\vec \xi\,)$ at a
given sign of $-\delta {\cal N}$ ($-$ or $+$ symbolizes rarefaction or
compression respectively).  We will call these stripes ``$-$stripes'' or
``$+$stripes''.

The energies $\varepsilon_{\pm} = |\varepsilon(\vec d_{\pm},\vec
\xi_{\pm})|$ are the quantities $P_{\pm}$ (see (\ref{identity}))
associated with $\pm$stripes formation. The corresponding $\mu_{\pm}$, as
follows from general thermodynamic considerations, are

\begin{equation}
\label{mu_P}
\tilde \varepsilon_{\pm} = u_0 + Q\varepsilon_{\pm}.
\end{equation}

\noindent Lest there be no contradiction with the fact of the coalescence,
energies $\tilde \varepsilon_{\pm}$ and $\epsilon_\pm$ are bound to
satisfy inequalities

\begin{equation}
\label{inequalities}
\epsilon_- <\tilde \varepsilon_- < \tilde \varepsilon_+ < \epsilon_+\,.
\end{equation}
\noindent When $Q\gg 1$ and $v(r)$ goes to zero over distances $R\ll
r_{ee}\sim Q^{1/2}$, they follow from simple estimates. Taking into
account that $|\vec\xi_{\pm}|\sim a_0$, and, corespondingly, $|\vec
d_\pm\times \vec\xi_{\pm}|\sim Q^{1/2}\sigma_0$, from Eq.~(\ref{dE}) we
obtain:  $\varepsilon_{\pm}\sim (a_0Q^{1/2}/R)u_0$.  On the other hand,
$|\epsilon_-| \sim u_0\sim v(r_{ee})$, and hence, $\tilde \varepsilon_-
\gg |\epsilon_-|$. In the case under consideration $\epsilon_+\sim
v(r_{min})$, where $r_{min}$ is the least of the distances between the IP
and the S-crystal sites. This energy is much more than $\varepsilon_+$ as
$R\ll r_{ee}$.

To make sure that the inequalities (\ref{inequalities}) hold for other
$v(r)$ and $R/r_{ee}$ we have computed $\varepsilon_{\pm}$ (basing on
Eq.~(\ref{dE}) and Eq.~(\ref{mu_P})) in parallel to the Monte-Carlo
computer studies of the coalescence. They have confirmed that the
inequalities are really the case for all $v(r)$ under consideration and
for all $Q$, maybe except $Q= 2$.

Together with the mutual repulsion of $\pm$stripes of the same sign the
inequalities (\ref{inequalities}) lead to the conclusion that $+$stripes
or $-$stripes do constitute the GS superstructure in the vicinity of
$1/Q$.  The position of each $\pm$ stripe -- a constituent of the
superstructure -- is determined by the stripe ``coordinate'', $l$, which
is the total number of particle lines (with the PTV $\vec d_{\pm}$)
between this stripe and some fixed one ($l= 0$). A set of these
coordinates determines uniquely the 2D FEP space structure. Therein lies
the LOD.

\paragraph{The GS superstructure of stripes.}
\label{superstructure}
The GS arrangement of the $\pm$stripes is governed by the pair potential
of the stripe-stripe interaction,

\begin{equation}
\label{vss}
V_{\text {ss}}^{\pm}(l)
=\sum\nolimits_{n=l+1}^\infty{\sum\nolimits_{\vec r}}^{\prime}u_{\vec
\xi_{\pm}, -\vec \xi_{\pm}}(\vec r - n \vec f_{\pm} -\vec
\xi_{\pm})
\end{equation}

\noindent where inter-stripe ``distance'' $l= 1,2,\ldots$; $\vec f_{\pm}$
is an S-crystal PTV other than $\vec d_{\pm}$,
$\Sigma^{\prime}_{\vec r}$ means the same as in Eq.~(\ref{dE})
($\vec d, \vec f =\vec d_{\pm}, \vec f_{\pm}$).  For all $v(r)$ under
consideration $\Sigma^{\prime}_{\vec r}(n) > 0$, and
$\Sigma^{\prime}_{\vec r}(n) >\Sigma^{\prime}_{\vec r}(n+1)$. Hence,
$V_{\text{ss}}(l)> 0$ is a convex function of $l$. This enables us to
describe the $\pm$stripes superstructure at $\vartheta - Q \neq 0$
($\vartheta = 1/\rho$) on the basis of the universal 1D algorithm
\cite{Hubbard,Bak,Synay}, considering the stripes as the "particles" of an
effective 1D FEP:

\begin{equation}
\label{super}
l_m = [m/c_{\pm}]; \quad c_{\pm} = |\vartheta - Q|/\sigma_{\pm}, \;
\sigma_{\pm}= |\vec d_{\pm}\times \vec\xi_{\pm}|
\end{equation}

\noindent where  $[\cdots]$ is the integral part of a number, $m$
enumerates the $\pm$stripes; integer $l_m$ is the coordinate of $m$-th
stripe, which is a pair of neighboring  lines of electrons
$\vec r_{m,1}(k)= k\vec d_{\pm} + l_m \vec f_{\pm} + m \vec \xi_{\pm}$
and $\vec r_{m,2}(k) = \vec r_{m,1}(k) + \vec f_{\pm} + \vec \xi_{\pm}$
($k = 0,\pm 1,\ldots$).
The superstructure described by Eq.~(\ref{super}) is thus a mixture of
$-$stripes ($\vartheta - Q> 0$) or $+$stripes ($\vartheta - Q <0$) and
unperturbed stripes of the S-crystal which are parallel to $\vec d_{\pm}$,
so that $c_{\pm}= N_s/{\cal N}_s$ is the concentration of the
$\pm$stripes; $N_s$ is their number; ${\cal N}_s$ is the total number of
the $\pm$stripes and the S-crystal ones. The number of unperturbed stripes
between $m$-th and $m+1$-th $\pm$stripes equals $l_{m+1} - l_m -1$.

\paragraph{An algorithm for arrangement of electrons' lines}
\label{lines}
S-crystals with $\vec f_{\pm} = q_{\pm}\vec
\xi_{\pm}$, where $q_{\pm}= Q/\sigma_{\pm}$ is an integer are of frequent
occurrence. Particularly, this occurs of necessity for a triangular host
lattice (section \ref{THL}), and also for $\sigma_{\pm} =\sigma_0$, as is
typical of S-crystals on a host lattice of a lower symmetry. In such a
case  the above-mentioned electron lines of both types, $\vec r_{m,1}(k)$
and $\vec r_{m,2}(k)$, fall into the class of electron lines $k\vec
d_{\pm}+ l\vec \xi_{\pm}$ ($k=0,\pm 1,\ldots$; $l$ is an integer), which
can be considered as 1D ``particles'' with ``coordinates'' $l$. Their
arrangment, as follows from Eq.~(\ref{super}), obeys the algorithm:

$$l_m =[\overline s\,m], \quad \overline s =q_{\pm}\,\mp\, c_\pm,$$

\noindent where $l_m$ is the ``coordinate'' of the $m$-th line, $\overline
s$ is the mean line separation measured in units of $|\vec \xi_{\pm}|$.

\subsection{ Devil staircase.}
\label{Devil}
The dependence of $c_{\pm}$ (or $\rho$) on $\mu$, much the same to the 1D
FEP \cite{Hubbard,Bak,Synay}, is a well-developed fractal structure, a
devil staircase whose steps occur at all {\em rational} $c_{\pm}= M/L \leq
1$ ($M,L$ are coprime integers). At given $M,L$ the GS configuration of
the 2D FEP is thus a ``FEP crystal'' with $L$ electrons per cell and with
PTVs $\vec d_{\pm},\,L\vec f_{\pm} + M\vec \xi_{\pm}$.

In the commonly occuring case that $\vec f_{\pm}$ is a multiple of $\vec
\xi_{\pm}$ (section \ref{d}) the steps' widths, $\Delta\mu = \Delta\mu
(M/L)$, can be found by direct application of the 1D theory
\cite{Bak,Synay}, considering the energy of the line-line repulsion,
$${\cal V}(l)= \sum_{k=-\infty}^\infty v\,(|k\vec d_{\pm} + l\vec
\xi_{\pm}|)$$

\noindent ( $l$ is the distance between interacting lines), as the 1D pair
potential. This produces
$$\Delta\mu = {\cal L}\sum_{m=1}^{\infty} m \left({\cal V}({\cal
L}m-1)-2{\cal V}({\cal L}m)+ {\cal V}({\cal L}m+1)\right),$$

\noindent where ${\cal L} = q_{\pm}L \mp M$ is the period of the lines'
pattern. The expression in the brackets is positive since in the case
under consideration ${\cal V}(l)$ is a convex function. Generally,
$\Delta\mu (M/L)$ are expressed in terms of $V_{\text {ss}}^{\pm}(l)$ by a
slight modification of the 1D theory.

\subsection{ j-branches and first-order transitions in the ground state of
the 2D FEP.}
\label{branch} The algorithm (\ref{super}) can be extended over the whole
$c_\pm$ range, $0<c_\pm<1$, provided the crystal with one particle per
cell (``S$^\prime$-crystal'' with PTVs $\vec d_{\pm},\:\vec f_{\pm}+\vec
\xi_{\pm}$) which arises at $c_\pm= 1$ ($\vartheta = Q \pm \sigma_{\pm}$)
is stable (i.e. it is another S-crystal) or metastable.  This follows from
the fact that i/ owing to the coalscence of defectons macroscopically
small variations in $\vartheta$ generate, at any $c_\pm$, 1D defects only;
ii/ these 1D defects, according to our computer calculations, have the
same PTV, $\vec d_{\pm}$, for all $c_\pm$.

Moreover, due to (meta)stability of the S$^\prime$-crystal the algorithm
(\ref{super}) holds over a $\vartheta$ range adjacent to the interval $[ Q
-\sigma_-, Q + \sigma_+]$.  In such a case Eq.~(\ref{super}) determines a
mixture of stripes of new geometry which are characterized by a new triple
of vectors, $\vec d_\pm^{\,\prime},\:\vec f_\pm^{\,\prime},\: \vec
\xi_\pm^{\,\prime}$, the analogues of $\vec d_{\pm},\:\vec f_{\pm},\: \vec
\xi_{\pm}$, and the $\pm$stripes concentation $c_\pm^{\prime}= |\vartheta
- Q\pm \sigma_{\mp}|/|\vec d_\pm^{\,\prime}\times \vec
\xi_\pm^{\,\prime}|$.  Transition from one geometry to another is
continuous in $\vartheta$ since $c_\pm^{\prime}$ goes to zero when
$\vartheta \rightarrow Q + \sigma_{\pm}$.

Continuously extending the algorithm (\ref{super}) in the manner shown
above, we obtain the ``$j$-branch'' (we introduce the index $j$ again)
which comprises all (meta)stable structures Eq.~(\ref{super}) connected in
continuity with the starting S-crystal. The corresponding energy,
$E_j(\vartheta)$, can be easily found in terms of $V_{\text
{ss}}^{\pm}(l)$, using Eq.~(\ref{super}). As a rule, there exist different
S-crystals belonging to the same $j$-branch.  On the other hand, as we
have computed, intersections of different $E_j(\vartheta)$, and hence,
{\em zero-temperature first-order transitions in variables $\mu$ or $P$}
(a type of polymorphism), are universally present in the 2D FEP.  (See
example in section \ref{THL}). The dependence of $E_g$ on $\vartheta$ is
the function which comprises all stable portions of all $E_j(\vartheta)$.

Thus, owing to the LOD described above the GS of the 2D FEP is fully
determined by the S-crystals PTVs, $m^j_{\kappa\lambda}$, the
``directors'', $\vec d_{\pm}^{\,j}$, and the displacement vectors, $\vec
\xi_{\pm}^{\,j}$, together with the set of $E_j(\vartheta)$ intersection
points which are the only GS characteristics changing on small variations
in $v(r)$. All these quantities can be computed on the basis of
Eq.~(\ref{dE}) and Eq.~(\ref{super}) by a self-consistent procedure,
finding the S-crystals together with the $j$-branches.  We have found the
GS for triangular and square host lattices as well as for a number of
those with central symmetry only. The computation has not revealed
principal differences between GS properties of 2D FEP with different
geometry of host lattices, at least for those which are not significantly
anisotropic.

\subsection{Example.}
\label{THL}
Here we illustrate the above general results with a triangular host
lattice (THL).  All triangular lattices on the THL are necessarily
S-crystals. This follows from the fact that it is the triangular lattice
that realizes the {\em absolute} energy minimum of the system whose
electrons are free to move.  Such S-crystals are ``$p,q$-crystals'' with
PTVs $p\vec a_1 + q \vec a_2, \ p\vec a_2 + q \vec a_3$ and $\vartheta =
p^2 + q^2 - pq$ ($p,q$ are arbitrary integers, $\vec a_{1,2,3}$ is a
triple of THL PTVs which are equal in the modulus and form an angle of
$120^{\circ}$ with each other).  Using the procedure discussed in section
{\ref{branch}}, we have found that all $0,q$-crystals belong to the same
$j$-branch (the main branch), which covers the range $4\leq \vartheta
<\infty$. The S$^{\prime}$-crystals of the $0,q$-ones are S-crystals too.
They occur at $\vartheta = q(q+1)$ ($2\leq q <\infty$) and have PTVs
$q\vec a_\kappa, \,(q+1)\vec a_\lambda$ \ ($\kappa,\lambda= 1,2,3$;
$\kappa \neq \lambda$). The stripe structures (\ref{super}) have the same
PTV, $q\vec a_\kappa$, for all $\vartheta$ of the interval $[q(q-1),
q(q+1)]$, their $\vec \xi_{\pm}$ being $\pm \vec a_\lambda$ ($\kappa\neq
\lambda$).

When $p,q \neq 0$, $j$-branches of different $p,q$-crystals are distinct.
They do not have mutual intersections, but all intersect the main branch,
the intersections occuring at rather small concentrations of the
$p,q$-crystals' $\pm$stripes. In other words, the intervals of
$p,q$-crystals stability ($p,q \neq 0$), and correspondingly main-branch
metastability, turn out to be narrow.

\subsection{The limit of $\gamma\gg 1$.}
\label{bigR}
So far the case of $\gamma\lesssim 1$ (section {\ref{defecton}}) has been
discussed. Here we outline the limiting case $\gamma \gg 1$. It is
realized when the Coulomb interelectron forces are screened by conductors
which are at distances $\gg r_{ee}$ from the 2D FEP. Modelling such a
situation by the potential $v(r)\propto r^{-1}\exp(-r/R)$ with $R\gg
r_{ee}$, we have computed that the energies $\epsilon_\pm,
\;\varepsilon_\pm$ satisfy the inequalities $\varepsilon_-< \epsilon_-
<\epsilon_+ < \varepsilon_+$, which are opposite to those of
Eq.(\ref{inequalities}). Due to this fact it is separated zero-dimensional
defects of the S-crystal that form the GS superstructure for $\rho$ which
are sufficiently close to $1/Q$. We have revealed that these
zero-dimensional defects are ``bidefectons'', which are complexes
consisting of two bound defectons.  Well-separated bidefectons can be
considered as new particles on the S-crystal as the host lattice, the mean
bidefecton separation, $r_d$, being equal to $\sim |2(\rho -
1/Q)|^{-1/2}$. The effective pair potential of a mutual bidefecton
repulsion is characterized by the same space parameter, $R$, as $v(r)$.
If $r_d \gtrsim R$, the bidefectons, according to the general results of
sections \ref{defecton}, \ref{d}, are bound to be ordered into stripes
arranged by the algorithm (\ref{super}). Extention of this reasoning to
the case of $R\gg r_d$ leads to new stripe-like superstructures consisting
of zero-dimensional defects of ``new'' S-crystals, and so on. Eventually a
well-developed fractal arises. Though details of its structure are still
to be determined, it is safe to say now that the LOD does take place for
$\gamma\gg 1$, too.

\subsection {Summary.}
\label{Summary}
\setcounter{paragraph}{0}
The above consideration shows that the electron ordering into stripes and
the effective lowering of dimension reside in the 2D FEP universally. In
essence, a combination of discreteness of electrons' positions with a
long-ranged interelectron repulsion is the only factor which gives rise to
this phenomenon. For this reason it is also bound to arise with an
external disorder present, the stripes being fractured and pinned by the
disorder. Thus, stripe formation in 2D and layered narrow-band conductors
can be considered to be the principle signature of a 2D FEP.

\paragraph{The charge ordering in cuprates as a manifestation of
a 2D FEP.}
\label{summary1}
From the above standpoint the charge ordering in $CuO_2$ planes of {\em
high-temperature superconductors} (cuprates) \cite{HighTc1}
(neutron scattering), \cite{HighTc2} (channeling) is of especial
interest.  The fact that it takes place even with very low dopping
\cite{HighTc2} suggests that a 2D FEP might be present in these systems
primordially. One can envision that formation of ionized oxygen molecules,
$O_2^{--}$, in oxygen planes gives a certain energy gain even in cuprates
of the {\em stoichiometric} composition \cite{footnote2}. In consequence,
a part of electrons leaves the oxygen planes for $s$-orbits of
$Cu^{++}$ ions in $CuO_2$ planes, resulting in formation of a number of
ions $Cu^+$.  Since the amplitude of electron hopping
$Cu^+ \leftrightarrow Cu^{++}$ is relatively small, the $Cu^+$ ensemble
should be expected to be a 2D FEP, the concentration of the $Cu^+$ and,
correspondingly, of the $O_2^{--}$ being determined by thermodynamic
equilibrium between the 2D FEP and the ensemble of the $O_2^{--}$. It is
evident that stripe formation in the 2D FEP of $Cu^+$ ions inevitably
brings to existence $O^-$ superstructures  in $CuO_2$ planes. Their PTVs
are likely to be the same as that of the $Cu^+$ FEP.

In the connection with the aforesaid it should be noted that a simple
explanation of the high-temperature superconductivity can be offered in
terms of the 2D FEP taking into account the finiteness of the bandwidth
\cite{FNT93_UFN95}. It lies in the fact that a virtual exchange of 2D FEP
elementary excitations between oxygen holes (which are known to be free
charge carriers in the dopped cuprates) leads inevitably to a mutual
effective attraction of the holes and thereby to superconductivity (of
purely Coulomb origin) with high $T_c$. Our preliminary studies have shown
that the lowest-energy elementary excitations in the cuprate 2D FEP are
kinks on the disorder-fractured stripes.

\paragraph{Some expectable  features of the 2D FEP
thermodynamics and conductivity as a consequence of the stripe formation.}
\label{summary2}
Our preliminary studies have shown that the effective lowering of
dimension in the ground state of the 2D FEP accounts for a fairly
interesting and unusual low-temperature thermodynamics. It is
characterized by first-order transitions in $T,\mu$-plane ($T$ is the
temperature) from the FEP crystals (section \ref{Devil}) slightly
perturbed by an ideal gas of separate defectons (they arise due to thermal
activation) to a strongly correlated liquid of
thermally fractured stripes (``FEP liquid'') where there is no
long-ranged order.  The melting temperature as the function of $\mu$ turns
out to be reduced to {\em zero} at the endpoints of the intervals of the
devil staircase.  Therefore, at any $T\neq 0$ there is a set of
alternating $\mu$ intervals which correspond to the FEP crystals or the
FEP liquid.

Conduction in the 2D FEP liquid is expected to be by movement of
kinks of the fractured stripes, each kink carrying a fractal charge
(measured in units of $e$). That in the FEP crystals is of the common
Drude type, the charge carriers being $\pm$defectons with the charge $\pm
e$. With a change in $\mu$ (at a fixed $T$) these conduction mechanims
alternate, resulting in pronounced 2D FEP resistivity oscillations which
reflect the ground-state devil-staircase dependence of $\rho$ on $\mu$:
the oscillations' peaks are bound to occur close to the {\em rational}
filling factors of the FEP crystals which survive at a given $T$. This
phenomenon is yet another distinctive mark of the 2D FEP. We have found it
to be very similar to the resitivity oscillations of a conductive sheet in
a system metal -- n-type semiconductor -- p-type semiconductor
\cite{Pepper}, which stil remain to be explained. We are going to
publish the results concerning this issue in the near future.

It is remarkable that an artificially created external perturbation
localized within a small region can block up conduction over all FEP
liquid, pinning only one stripe. The most appropriate systems to test this
experimentally are perhaps granular thin films like those described in
\cite{Osifchin}. A similar phenomenon was reported in \cite{Nejoh}. Yet
granular films used in these experimental studies were highly disordered,
and it is unclear now whether the above theory works in such a situation.

\paragraph*{Acknowledgments.}
We gratefully acknowledge discussions with M.~Pepper and P.~Wiegmann.

\begin{figure}
\caption{The coalescence for square (Fig.~1a, $Q=9$) and triangular
(Fig~1b, $Q=16$) host lattices ($\nu=-5$).  Here $\circ$ denotes
host-lattice sites, $\protect\large\bullet$ - particles; $\odot$ -
antiparticles; $\otimes$ - holes; $\rightarrow$ - dipoles; solid boxes
mark off a single defecton. The shells of dipoles surrounding holes are
marked off by the dash-dotted parallelograms. The dotted lines show
nucleation of the elementary stripes enumerated by $1,2,\ldots$. In the
case (a) boundary effects dominate the mutual repulsion of the unfinished
$-$stripes; in the case (b) the tendency to $-$stripes divergence is seen.
Both configurations refer to the pair potentials  which meet the condition
$\gamma \sim 1$.}

\label{fig1}
\end{figure}


\begin{references}
\bibitem{footnote1} The $\rho$ values are assumed to be not too close to
$\rho = 1$, so that we can neglect  the configurations with
$n(\vec r) = 2$ in (\ref{H}) and take no account of the spin variables.
\bibitem{footnote2} A point in favor of this fact is a
noticeable disorder ($\sim 20\%$) in arrangement of $O^-$ along c-axis of
$\protect La_{2-x}Sr_xCuO_4$ which was revealed (by channeling
method\cite{HighTc3}) even for low $x$.
\end{references}
\end{document}